\documentclass[11pt,a4paper]{article}

\usepackage{amssymb}
\usepackage{latexsym}
\usepackage{multirow,booktabs}
\usepackage{threeparttable}
\usepackage{graphicx}
\usepackage{psfrag,graphicx}
\usepackage{pst-text}
\usepackage{amsmath}
\usepackage[latin1]{inputenc}
\usepackage{rotating,booktabs}
\usepackage{appendix}
\usepackage{tkz-graph}
\usepackage{pstricks,pst-node,pst-tree}%
\usepackage{fancyhdr}

\def\mybibliography#1{{\begin{center} \bf References \end{center}}\list
 {}{\setlength{\leftmargin}{1em}\setlength{\labelsep}{0pt}
\itemindent=-\leftmargin}
 \def\newblock{\hskip .02em plus .20em minus -.07em}
 \sloppy\clubpenalty4000\widowpenalty4000
 \sfcode`\.=1000\relax}
\newbox\TempBox \newbox\TempBoxA
\def\uw#1{%
  \ifmmode\setbox\TempBox=\hbox{$#1$}\else\setbox\TempBox=\hbox{#1}\fi%
  \setbox\TempBoxA=\hbox to \wd\TempBox{\hss\char'176\hss}%
  \rlap{\copy\TempBox}\smash{\lower9pt\hbox{\copy\TempBoxA}}%
}
\newbox\TempBox \newbox\TempBoxA
\def\uwd#1{%
  \ifmmode\setbox\TempBox=\hbox{$#1$}\else\setbox\TempBox=\hbox{#1}\fi%
  \setbox\TempBoxA=\hbox to \wd\TempBox{\hss\char'176\hss}%
  \rlap{\copy\TempBox}\smash{\lower10pt\hbox{\copy\TempBoxA}}%
}
\def\mathunderaccent#1{\let\theaccent#1\mathpalette\putaccentunder}
\def\putaccentunder#1#2{\oalign{$#1#2$\crcr\hidewidth
\vbox to.2ex{\hbox{$#1\theaccent{}$}\vss}\hidewidth}}

\newcommand{\bee}{\begin{eqnarray*}}
\newcommand{\eee}{\end{eqnarray*}}
\newcommand{\be}{\begin{eqnarray}}
\newcommand{\ee}{\end{eqnarray}  }

\def \R{I \negthinspace \negthinspace R}

\newcommand{\bbeta}{\mbox{\boldmath $\beta$}}
\newcommand{\mmu}{\mbox{\boldmath $\mu$}}

\def\ni{\noindent}

\def\be{\begin{eqnarray*}}
\def\ee{\end{eqnarray*}}

\def\bbeta{\mbox{\boldmath $\beta$}}

\begin{document}

\thispagestyle{empty}
% below is our logo

%{\hbox{\footnotesize\rm Journal of Data Science {\footnotesize\bf
% Vol.}(2014), 143-155}\hfill}

%$\ $\par
\vspace{0.4pc}

% Title next
\begin{center}
{\large\bf HYBRID BAYESIAN ESTIMATION IN THE ADDITIVE HAZARDS MODEL}
\end{center}

% Authors next
\vspace{.2cm}
\begin{center}
\renewcommand{\thefootnote}{\fnsymbol{footnote}}

\'{A}lvarez\ Enrique Ernesto$^{1}$, Riddick\ Maximiliano Luis$^{2}$\\
~\\

{\it} $^{1}$Instituto de C\'{a}lculo, Universidad de Buenos Aires - CONICET\\ Universidad Nacional de Luj\'{a}n, Argentina \\
mail: enriqueealvarez@fibertel.com.ar\\
%\email{enriqueealvarez@fibertel.com.ar}     Department\ of\ Biomedical\ Informatics, Asia\ University\\
$^{2}$NUCOMPA and Departamendo de Matem\'{a}tica, Facultad de Ciencias Exactas, 
Universidad Nacional del Centro de la Provincia de Buenos Aires, Tandil (7000), Argentina,\\
mail: mriddick@nucompa.exa.unicen.edu.ar\\

\end{center}

%\linenumbers

{\small
	\begin{quotation}
		\ni {\it Abstract}:%~~Suppose we have a sample of n individuals who may experience a certain event of interest over a window $[0,\tau]$. We denote by $T^*_i$ the true, possibly unobserved, time to occurrence for the $i$-th individual. Because some individuals experience censoring at times $C_i$, their duration until the event is observed only when $C_i \geq T^*_i$. In classical Survival Analysis, it is of interest to study the 
		%sojourn times as related to observed individual covariates, which we assume time-independent and denote by $Z_i$.  In the literature, models for survival data typically focus on the so-called hazard rate, which we assume takes the additive form $\lambda(t,\beta)=\lambda_0(t) + z'\beta$ due to Aalen (1980), where $\lambda_0(.)$ is the \textit{baseline hazard function} and $\beta$ is a vector of coefficients. Alternative approaches abound in the literature, the most common being Cox's (1972) proportional hazards model and the Accelerated failure time model.
		
		Hereby we propose a Bayesian method of estimation for the semiparametric Additive Hazards Model (AHM) from Survival Analysis under right-censoring. With this aim, we review the AHM revisiting the likelihood function, so as to comment on the challenges posed by Bayesian estimation from the full likelihood. Through an algorithmic reformulation of that likelihood, we present an alternative method based on a hybrid Bayesian treatment that exploits Lin and Ying (1994) estimating equation approach and which chooses tractable priors for the parameters. We obtain the estimators from the posterior distributions in closed form, we perform a small simulation experiment, and lastly, we illustrate our method with the classical Nickels Miners dataset and a brief simulation experiment.

		\vspace{0.4cm}
		% Key words next, notice the style.
		\ni {\it Keywords}:~~Additive Hazards Model, Survival Analysis, Bayesian Inference.
	\end{quotation}
}

\section{Introduction}
Suppose we have a sample of $n$ individuals who may experience a certain event of interest such as death, definite illness remission or permanent retirement from the labor force 
over an observation window $[0,\tau]$. We denote by $T^*_i$ the true, possibly unobserved, 
time to occurrence for the
$i$-th individual. Because some individuals experience censoring at times
$C_i$, their duration until the event is observed only when $C_i \geq T_i^*$. 
In classical Survival Analysis, 
it is of interest to study this variables as related to observed individual level characteristics, as measured by some independent variables or, frequenty called, \textit{covariates}. Notice that we are only considering that we are able to observe at most one event time for each individual. Even though this model could be extended for repeated events, we leave that as matter for future research. For example, in Medicine, it could be of interest how remission make depend on the choice of different drugs, or other individual specific characteristics. \\

In the literature, models for survival data typically are defined according to the so-called \textit{hazard rate},
%
%\[
 $\lambda_i(t,\theta):= \lim_{\varepsilon \downarrow 0}
\varepsilon^ {-1}\text{P}_\theta(T^*_i\leq t+\varepsilon|T^*_i > t),
$
%\]
which is a heuristic measure of the instantaneous risk at any given time.
Approaches for parametric models, where $\theta \in \Theta \subset \R^k$, are 
plenty. Some common choices are the exponential, Weibull or Gamma distributions (e.g., Lawless 2003). Alternatively, useful models have been proposed within the semi-parametric framework, in which the hazard function is decomposed in a nonparametric \textit{baseline} hazard function $\lambda_0(\cdot)$ and an Euclidean parameter $\bbeta \in \R^k$. In this setting the three most common models are:
(\textit{i})
%\begin{description}
%\item[
\textit{Cox's Proportional Hazards Model} (CPM), which decomposes the hazard multiplicatively proposing 
\begin{equation}
\lambda(t,\bbeta) = \lambda_0(t) \exp{(z' \bbeta)};  
%\label{eqn:phm}
\end{equation}
(\textit{ii}) the \textit{Additive Hazards Model} (AHM), which proposes instead 
an additive decomposition  
\begin{equation}
\lambda(t,\bbeta) = \lambda_0(t) + z' \bbeta,  
\label{eqn:ahm}
\end{equation}
and (\textit{iii}) the 
%
%\item[
\textit{Accelerated Failure Time Model} (AFTM), which takes a slightly different approach 
than the 
previous ones by 
proposing a rescaling for the duration times themselves. I.e., 
%$T^* \sim F_{\bbeta}(\cdot)$ where
%
%\[
$
T^* = T^*_0 \exp({Z'\bbeta)},
$
% \qquad \text{
where
%} \qquad
$
T^*_0 \sim F_0(\cdot)
$
% \quad \text{
is the baseline cumulative distribution function.
%}.
%\]
%
Interestingly, for the corresponding hazard functions, this entails that
\begin{equation}
\lambda(t,\bbeta) = \lambda_0 \left[t \, \exp(Z' \bbeta)\right] \exp(Z' \bbeta),
\label{eqn:aftm}
\end{equation}
%
%\end{description}
%
%Idea for future research. Is there any extension of the work of ?? that also 
%encompasses the AFM?
%
which is neither multiplicative nor additive.

Developments in Classical Estimation for the above three  models have been fairly vast. 
An extensive account for them can be found in some classical  
textbooks such as Kalbfleisch and Prentice (1980),  Klein and Melvin (2006) or Lawless (2003), among others.
As for Bayesian estimation, most of the statistical focus in the literature was
put on the Multiplicative Hazards Model, an extensive account of which can be found in the 
textbook of Ibrahim, Chen and Sinha (2001). Comparatively, Bayesian developments for the Additive 
Hazards Model and for the Accelerated Failure time Model have been fewer.  For the AHM, a 
recent literature review is provided by Alvarez and Riddick (2019), and for the AFT model, 
the relevant references are discussed in Zhang and  Lawson (2011). 

It is our goal in this study to propose a Bayesian method of estimation for the semi-parametric 
Additive Hazards Model under right-censoring. 
This paper is organized as follows. In Section \ref{sec:model} we review the 
Additive Hazards Model, we introduce the 
likelihood function and we discuss the challenges posed by either Classical or Bayesian estimation from the full likelihood.
We further discuss alternative approaches based on Cox's \textit{Partial Likelihood} and on 
the seminal work of Lin and Ying (1994), who proposed an approach which is based on an estimating 
equation method developed from Counting Process theory.

Section \ref{sec:hybrid} is our main contribution in this manuscript in which, based on a convenient expression of the likelihood, we present a new hybrid Bayesian 
method of estimation. This method combines the classical Lin and Ying's estimating 
equation with Bayesian priors for the Euclidean parameter and for the baseline hazard function.
We obtain 
the posterior distributions and provide formulas for the estimators in closed form. 
The availability of those closed-form expressions is one of the main advantages of 
our method, which avoids the necessity of relying on approximate sampling from the
posterior distributions or on other types of numerical approximation.
Section \ref{sec:simul} contains a small simulation experiment
that illustrates some of the properties of our hybrid Bayesian estimators. In
Section \ref{sec:data} we carry out a comparative Bayesian analysis for the 
Welsh Nickels Refiners dataset first introduced by Doll, Morgan and  Speizer (1970) and subsequently 
analyzed from a Classical perspective by Breslow and Day (1987),  Lin and Ying (1994), 
and Alvarez and Ferrario (2016), among others. Lastly, in section \ref{sec:conc}, we highlight the main contributions of this article, which is part of Maximiliano Riddick PhD thesis (2020).

\section{Additive Hazards Model}\label{sec:model}

The parameter vector $\bbeta$ is non-negative and $k$-dimensional, and 
$\lambda_0(\cdot) \in \mathfrak{L}$ is a \textit{baseline} hazard function. This entails that, denoting
\begin{equation*}
\mathfrak{L}:=\left\{h:\mathbb{R}^+ \mapsto \mathbb{R^+}, \quad \int_0^\infty h(s) ds =
\infty \right\},
\end{equation*}
the \textit{parameter space} for $(\bbeta,\lambda_0)$ 
is $\Theta = \mathbb{R}^{k+} \times \mathfrak{L}$. It is in that sense that the model is
\textit{semi-parametric}. It is noteworthy that in the general AHM formulation what needs to be positive is the hazard function itself, but not necessarily all the coefficients. In our context, there are two ways to accomplish that goal: ($\textit{i}$) by adding the specific constraint that the estimated $\hat{\lambda}(t) = \hat{\lambda}_0(t)+\beta'z$ be nonnegative and performing constrained inference; or ($\textit{ii}$) by using only positive covariates $z_k$, either as given by nature or constructing them purposely to achieve positivity, and forcing all the coefficients $\beta_k$ also to be positive. Althought it seems restrictive, this option could be extended to allow for negative parameters through the model proposed by Dunson \& Herring (2005). Our choice, which is sufficient yet not necessary to obtain nonnegative hazards, is the alternative that we adopted in this manuscript. The main advantage of being willing to live with this limitation is the availability of estimators in closed form. 
According to the hazard function of the model \ref{eqn:ahm}, and under right censoring, our interest is to estimate the parameters of the model. Under right censoring $\delta_i=1(t_i \geq u)$, the likelihood function in survival models is, calling $\Theta$ to the model parameters, \[ \mathcal{L}_n(\Theta)= \prod_{i=1}^n \lambda(t_i\vert \Theta) S(t_i\vert \Theta)^{\delta_i}. \]
In the next of this article, and for practicality, we will omit in the expression the conditional respect to the parameters of the model. So, in the AHM, the likelihood expression when $n$ i.i.d.\ triplets 
$(t_i,z_i,\delta_i)$ are observed results in 

\begin{eqnarray}
\mathcal{L}_n(\bbeta,\lambda_0) &=& 
\prod_{i=1}^n \left[f(t_i)\right]^{\delta_i}\left[S(t_i)\right]^{1-\delta_i}
=
\prod_{i=1}^n \left[\lambda(t_i)\right]^{\delta_i} S(t_i)
\nonumber
\\
&=&
\prod_{i=1}^n \left[\lambda_0(t_i)+\bbeta' z_i\right]^{\delta_i}
\exp\{-\Lambda_0(t_i)- t_i \, \bbeta' z_i  \}
,
\end{eqnarray}
since $S(t)=\exp\{-\Lambda(t)\}=\exp\{-\int_0^t \lambda(u)du\}$, where $\Lambda_0(t)$ represents the denoted \textit{cumulative baseline hazard function}, i.e., $\Lambda_0(t)=\int_0^t \lambda_0(u)du$. We will expand about this in the algorithm explanation.
\paragraph{Piecewise constant model.}
For simplicity, we will in this Section  
propose to model the 
baseline hazard function as independent of the Euclidean parameter and as a piecewise
constant function. E.g., we fix $m \in \mathbb{N}$ and we choose a partition 
$0=:s_0 < s_1 < \ldots < s_{m-1} < \infty =: s_m$. Define further a random nonnegative stepwise function
\begin{equation}
\mathbf{L}_0(t) := 
\begin{cases}
A_1 & 0 \leq t < s_1, \\
A_2 & s_1 \leq t < s_2, \\
\vdots \\
A_{m-1}  & s_{m-2}  \leq t < s_{m-1}, \\
A_m & t \geq s_{m-1}, 
\end{cases}
\label{eqn:steplambda}
\end{equation}
and let us call $\mathbf{A}$ the $m$-dimensional random vector with entries   
$A_1, \ldots, A_{m-1} \geq 0$, and with $A_m > 0$ (where the last entry is 
strictly positive in order to guarantee that any realization 
of the hazard function integrates to infinity). Strictly speaking, in a Bayesian setting, the random function  
$\mathbf{L}_0$ has distribution $\mathcal{Q}$, which is a probability measure on the space
$\mathfrak{L}$.
Several choices are possible for $\mathcal{Q}$. But, for simplicity, we assume throughout this manuscript
that the grid 
$s_0, \ldots, s_m$ is fixed and it was chosen before collecting any data.
\paragraph{Polynomial expansion of the Likelihood under the piecewise constant hazard model.}
The next construction is crucial since it makes available the Bayesian treatment developed in the next Section. 
First, we reconsider the likelihood including the piecewise exponential model proposed to the baseline hazard:
\begin{equation*}
\mathcal{L}_n(\bbeta,\lambda_0) = 
\exp\left\{-\bbeta' \sum_{i=1}^n t_i \, z_i  \right\} 
\exp\left\{ -\sum_{i=1}^n \Lambda_0(t_i) \right\} 
\prod_{i=1}^n \left[\lambda_0(t_i)+\bbeta' z_i\right]^{\delta_i}
.
\end{equation*}
Now, the trick is to rewrite the expression $\prod_{i=1}^n\left[\lambda_0(t_i)+\bbeta' z_i\right]^{\delta_i}$ according to the grid and the values defined in the piecewise exponential model. Let $n_{j}$ the number of observations that belong to each grid interval,
and calling  $(z_{k_{j}}, \delta_{k_{j}})$ to their corresponding covariate and censoring
indicator, this yields
\begin{align}
\mathcal{L}_{n}(\bbeta, \lambda_{0})
%&= 
%\exp\left\lbrace -\bbeta' \sum_{i=1}^{n}t_{i}z_{i} \right\rbrace 
%\prod_{i=1}^{n}\left[ \lambda_{0}(t_{i})+ \bbeta' z_{i} \right]^{\delta_{i}} 
%\exp\left\lbrace -\Lambda_{0}(t_{i}) \right\rbrace 
%\\
&=\exp\left\lbrace -\bbeta' \sum_{i=1}^{n}t_{i}z_{i}\right\rbrace 
\exp\left\lbrace -\sum_{i=1}^{n} \Lambda_{0}(t_{i}) \right\rbrace 
\prod_{j=1}^{m} \prod_{k_{j}=1}^{n_{j}}\left[ a_j+ \bbeta' z_{k_{j}} \right]^{\delta_{k_{j}}} 
\label{eqn:likexp1}
\\
%&=:
%f_{\mathcal{X}|\mathbf{L}_0,\mathbf{B}}
%\big{(}\{(t_i,z_i,\delta_i)\}:i=1,\ldots,n
%\big{)},\nonumber
\end{align}
%
%where the last notation emphasizes its Bayesian meaning as a joint density 
%of the sample given 
%the parameters under uniform (improper) priors.
%%  Then, calling $\pi(\mathbf{L}_0)$ the prior for the baseline,
%%  \begin{equation} 
%% f_{\mathbf{L}_0\vert \mathbf{B}, \mathcal{X}}
%% \propto
%% \exp \left \lbrace - \sum_{i=1}^{n} \Lambda_{0}(t_{i}) \right\rbrace 
%% \prod_{j=1}^{m} \left( \prod_{k_{j}=1}^{n_{j}}
%% \left[ a_j+ \bbeta' z_{k_{j}} \right]^{\delta_{k_{j}}}\right)\times \pi(\mathbf{L}_0) .
%% \label{eqn:margpostl}
%% \end{equation}
%
Notice that the formula in Equation (\ref{eqn:likexp1}) above does not correspond to any standard density
for which moment or quantile formulae, 
nor simulation algorithms, are readily available. Because of that, the previous statistical inference approaches to this model were based on numerical methods.\\
We now endeavor to find approximations to its first two moments, from a tractable re-expression of this formulae.
Let us expand the product 
\begin{equation}
\prod_{k_{j}=1}^{n_{j}}\left[ a_{j}+ \bbeta' z_{k_{j}} \right]^{\delta_{k_{j}}}
=
d_{0} + d_{1}a_{j} + \ldots 
+ d_{N_{j}}a_{j}^{N_{j}}, 
\label{eqn:polexp}
\end{equation}
which is  a polynomial of order $N_{j}=\sum_{k_{j}=1}^{n_{j}}\delta_{k_j}$, with 
coefficients $d_{i}$'s that depend on the sample 
$(t_{i},z_{i},\delta_{i})$, ${i=1,...,n}$ and on the parameters.
To estimate the coefficients of Equation (\ref{eqn:polexp}), Chernoukhov (2013, 2018) proposed a method
which essentially consisted in two steps: (\textit{i}) evaluating the polynomial at $(N_j+1)$ 
different points; and (\textit{ii}) solving a system of linear equations.
The drawback of that method is its numerical unstability.
Instead, in this manuscript we propose an alternative approach which is more
efficient numerically and which allows for estimators in closed-form. 
With that aim, we basically exploit 
a recursive relationship 
obtained for the polynomial coefficients. We explain this as follows.

\paragraph{Recursion}
Let us propose the following recursive method, based on how the polynomial coefficients changes when a degree $n$ polynomial is multiplied by a monomial $(x+b)$. 
If the number of uncensored times is $n$ (i.e., a degree $n$ polynomial, for a generic variable $\lambda$),
we denote the polynomial 
%
%\[ 
$P_{n}(\lambda)= \sum_{j=0}^{n}d_{j}\lambda^{j}
$.
% \]
%
When the sample is augmented by one uncensored observation, 
it is equivalently to multiply $P_{n}$ by $(\lambda+ \bbeta'z_{n+1})$, i.e., 
\begin{equation*}
P_{n+1}(\lambda) = \left( \sum_{j=0}^{n}d_{j}\lambda^{j}\right)(\lambda + \bbeta' z_{n+1}) 
=\sum_{j=0}^{n}d_{j}\lambda^{j+1} + \sum_{j=0}^{n}d_{j}\bbeta'z_{n+1}\lambda^{j}.
\end{equation*}
Then, the coefficients of $P_{n+1}$ will be updated as (\textit{i}) 
$d_{0}^{(n+1)}=d_{0}^{(n)} \,\bbeta'z_{n+1}$,
(\textit{ii})
$d_{j}^{(n+1)}=d_{j-1}^{(n)}+d_{j}^{(n)}\bbeta'z_{n+1}$, for $j \in \{1,...,n\}$,
and lastly 
(\textit{iii}) 
$d_{n+1}^{(n+1)}=d_{n}^{(n)}=1$.
With this recursive relationship between the coefficients, we can calculate the $d_{i}$'s in an
fairly easy and numerically efficient way.
According to this modification, we propose to estimate $\bbeta$, and then estimate $\lambda_0$. In order to achieve that goal, we deal with the likelihood expression $\mathcal{L}_n(\lambda_0\vert \bbeta)$ which becomes
\begin{equation}
\mathcal{L}_n(\lambda_0\vert \bbeta) \propto  
\prod_{j=1}^{m} \left( \sum_{k=0}^{N_{j}}d_{k}a_{j}^{k}\right)\exp\left\lbrace -\sum_{i=1}^n\Lambda_{0}(t_{i}) \right\rbrace . 
\label{eqn:likexp2}
\end{equation}
We endeavour now to obtain a tractable expression for 
$\exp\left\lbrace -\sum_{i=1}^n\Lambda_{0}(t_{i}) \right\rbrace
$
in  Equation (\ref{eqn:likexp2}), for each interval in the grid, expressing each $\Lambda_0(t_i)$ as function of each $a_j$.  For that, notice that, for each $a_{j}'s$, $\Lambda_0(t)$ (i) does not depend of $a_j$ if $t\leq s_{j-1}$, (ii) depends of $a_j$ and $t$ if $s_{j-1}<t\leq s_j$, or (iii)   
\begin{equation*}
\Lambda_0(t_i) := 
\begin{cases}
\Lambda_0(t_i) & t_i \leq s_{j-1}, \\
\Lambda_0(s_{j-1})+a_j(t_i-s_{j-1}) & s_{j-1} < t_i \leq s_j, \\
\Lambda_0(s_{j-1}) + a_j(s_j - s_{j-1}) + \int_{s_j}^{t_i}\lambda_0(u)du  & s_j < t_i. \\
\end{cases}
%\label{eqn:steplambda}
\end{equation*}
Then, calling $m_j= \# \{t_i > s_j \}$, after standard calculations 
we obtain 
\begin{equation}
\exp \left\lbrace -\sum_{i=1}^{n}\Lambda_0(t_i) \right\rbrace 
\propto \exp\left\lbrace-a_j\left(\sum_{k_j=1}^{n_j}(t_{k_j}- s_{j-1})+ m_j(s_j - s_{j-1})\right)\right\rbrace. 
\label{eqn:expexp}
\end{equation}
Therefore, combining Equations (\ref{eqn:likexp2}) and (\ref{eqn:expexp}) 
we see that $\mathcal{L}_n(\lambda_0\vert \bbeta)$ is proportional to a mixture of Gamma distributions,
i.e., 
\begin{eqnarray}
\label{eqn:CHGP}
\mathcal{L}_n(\lambda_0\vert \bbeta)
 \propto
\prod_{j=1}^{m} \left(\sum_{k=0}^{N_{j}}d_{k}a_{j}^{k}
\exp \left\{\!\!-a_j\left(\sum_{k_j=1}^{n_j}(t_{k_j}\!-\! s_{j-1})+ m_j(s_j\! -\! s_{j-1})\right)
\right\}\right). 
\end{eqnarray}
Being a mixture of Gammas, the display above provides a tractable expression which we extensively exploit in the sequel.

Without this convenient expression, several algorithms were presented in literature with innovative adaptations of acceptance rejection sampling,
Metropolis-Hastings
and Metropolis within Gibbs (e.g., Turkmann, Paulino \& M\"uller 2020). Instead, our formulation, because it leads to a mixture of Gamma distributions, enables avoiding the Gibbs sampler and provide estimators in closed form.

\subsection{Uninformative Priors}
According to covariates and censoring indicators, under the piecewise constant hazard model for the baseline hazard we arrive to mixture of Gamma distributions expression to the likelihood function for the AHM. We leave analysis under non-informative priors formulations as Laplace prior (equals to $1$ over the parametric space), or MAXENT priors for a future research. In the present article the objective is to develop Bayesian inference according to informative priors.

\subsection{Informative Priors}
\label{subsec:Inf Prior}
As mentioned above, in a Bayesian treatment, we view the parameters as realizations of a random vector
$\mathbf{B}=\bbeta$ and a \textit{random curve} $\mathbf{L_0}=\lambda_0(\cdot)$. In this Subsection, we 
attempt to specify informative \textit{prior} distributions in the following manner:
\begin{description}
	\item[Prior for the Euclidean Parameter.] Consider
	\begin{equation}
	\mathbf{B} \sim N_k(\mmu_\beta,\mathbf{C}_\beta)|_{\mathbb{R}^{k+}},
	\label{eqn:prioribeta}
	\end{equation}
	this is a truncation of a multivariate normal distribution with mean vector $\mmu_\beta$ and positive definite covariance matrix  $\mathbf{C}_\beta$, to the positive
	orthant $\mathbb{R}^{k+}$. This entails that the density function for $\mathbf{B}$ 
	is given by 
	\begin{equation}
	f_\mathbf{B}(\bbeta)=
	\dfrac
	{\displaystyle N_{\beta}^k
		%\left(\frac{1}{\sqrt{2 \pi}}\right)^k %\,
		 %|\det(\mathbf{C_\beta})|^{-\frac{1}{2}}
		%\exp\left\{\!-\frac{1}{2}(\bbeta\!-\!\mmu_\beta)' \mathbf{C_\beta}^{-1} %(\bbeta\!-\!\mmu_\beta)\right\}
		\prod_{j=1}^k I(\beta_j \geq 0)
	}
	{
		\displaystyle\int_0^\infty\!\!\! \ldots \int_0^\infty \! N_{\beta}^k %\left(\frac{1}{\sqrt{2 \pi}}\right)^k  |\det(\mathbf{C_\beta})|^{-\frac{1}{2}}
		%\exp\left\{\!-\frac{1}{2}(\bbeta\!-\!\mmu_\beta)' \mathbf{C_\beta}^{-1} %(\bbeta\!-\!\mmu_\beta)\right\}
		d\bbeta.
	}
	\label{eqn:priorbetadens}
	\end{equation}
	where $N_{\beta}^k=\left(\frac{1}{\sqrt{2 \pi}}\right)^k %\,
	|\det(\mathbf{C_\beta})|^{-\frac{1}{2}}
	\exp\left\{\!-\frac{1}{2}(\bbeta\!-\!\mmu_\beta)' \mathbf{C_\beta}^{-1} (\bbeta\!-\!\mmu_\beta)\right\}
	$. 
	The denominator is the probability P$(\beta_1 \geq 0; \ldots; \beta_k \geq 0)$ under 
	the nontruncated normal distribution. It is a constant that we denote by 
	$\Omega(\mmu_\beta, \mathbf{C_\beta})$ and which depends only on $\mmu_\beta$ and $\mathbf{C_\beta}$. Apart for its mathematical tractability, an additional advantage of choosing a prior based on a normal distribution is that it may simplify the process of prior elicitation. This is because field experts are often reasonably acquainted with them, by specifying the mean, mode or some of its quantiles. 
	% It could be calculated using conditional normal distributions by observing that if 
	%% $\mathbf{B} \sim N_k(\mmu_\beta,\digma\mathbf{C}_\beta)$ (nontruncated), 
	%% then $\mathbf{W}:=\mathbf{C_\beta}^{-1/2} (\bbeta-\mmu_\beta)$ is a vector of $k$ independent 
	%% standard normal variables. Therefore, standard calculations using the Jacobian method 
	%% lead to
	%% %
	%% \begin{eqnarray}
	%% \Omega(\mmu_\beta, \mathbf{C_\beta})
	%% &=&
	%% \text{P}(\beta_1 \geq 0; \ldots; \beta_k \geq 0)
	%% \\
	%% &=&
	%% \text{P}(\beta_1 - \mu_{\beta_1} \geq - \mu_{\beta_1}; \ldots; \beta_k - \mu_{\beta_k} \geq 
	%% - \mu_{\beta_k})
	%% \end{eqnarray}

	\item[Prior for the cumulative baseline hazard function.]  
	In order to making available certain dependence among the baseline hazard function parameters,
	we opted to model it as a Gamma process, with a pre-specified increasing and left continuous function $\alpha(t)$ and a scale parameter $c$. 
	So, calling $\eta_i:=\alpha(s_i)-\alpha(s_{i-1})$ the increment of the function $\alpha(t)$ in the interval $[s_{i-1},s_i)$, the joint prior density for the cumulative baseline parameters $\Lambda_{0_j}^+$ according to the time grid selected is 
	\begin{equation}
	f_{\mathbf{\Lambda_{0_j}^+}}(a_{0_1}^+,\ldots,a_{0_m}^+)=\prod_{i=1}^m \dfrac{c^{\eta_i}}{\Gamma(\eta_i)}
	(a_{0_i}^+)^{\eta_i-1}\exp\left(-c \, a_{0_i}^+\right),
	\label{eqn:infpostl}
	\end{equation}
where the introduced new parameters $(a_{0_1}^+,\ldots,a_{0_m}^+)$ are a realization of the random vector $\Lambda_0^+ = (\Lambda_{0_1}^+,\Lambda_{0_2}^+,\cdots,\Lambda_{0_m}^+)$, and are constructed according to the baseline hazard increments. We provide algebraic details of this calculations below in Section \ref{SS:EBH}.
We let $\mmu_\beta=\mathbf{1}_k$ be a vector of ones, and $\mathbf{C_\beta}=\omega \, \mathbb{I}_k$,
i.e., a constant $\omega>0$ times the identity matrix of order $k$.
Alternative methods which rely on choosing the hyperparameters for both $\mathbf{B}$ and $\mathbf{L}_0$ 
under prior elicitation from expert knowledge are currently under research.
\end{description}

\section{The Hybrid Bayesian Method}
\label{sec:hybrid}
We attempt to develop a Bayesian method that achieves two goals,
(\textit{i}) it disentangles estimation of $\bbeta$
from the baseline hazard function $\lambda_0(\cdot)$, as the latter is often treated as a nuisance in many applications, and (\textit{ii}) it generates 
estimators in 
closed form. Those are two important goals because %
Survival Analysis typically has two objectives, namely: (\textit{i}) prediction of the survival 
of an individual with given covariate values, and/or (\textit{ii}) assessing how survival
or the risk of the event occurring at any given time may depend on some covariates. While 
for the first objective (i.e., prediction)  estimation of both the baseline hazard function
and the Euclidean parameters is necessary, for the second objective only the Euclidean 
parameters are needed. As already mentioned, 
straight application of the maximum likelihood approach for 
either of the three models mentioned in this article, namely PHM, AHM or AFTM, estimates the baseline hazard function and the Euclidean parameter jointly 
(e.g., Andersen \textit{et.al.}, 1993) has
presented some computational difficulties that encouraged researchers to propose 
alternative estimation methods in the literature. For instance, for the PHM, Cox has 
developed the approach of the so-called \textit{Partial Likelihood}, which made it possible
to estimate $\bbeta$ disentangled from $\lambda_0(\cdot)$. First, we propose an approach to estimate $\bbeta$, and then we develop a method to estimate the baseline hazard given the estimation of $\bbeta$.
\subsection{Estimation of $\bbeta$}
In the AHM context, a classical way to estimate 
$\bbeta$ uses an estimating equation approach instead of the score function from either 
maximum likelihood or partial maximum likelihood. This is because neither of them allows the estimation
of $\bbeta$ to be disentangled from estimation of $\lambda_0$ and, furthermore, they do not
allow estimators in closed form. The pioneering work of Lin and Ying (1994) proposed 
an estimating equation for $\bbeta$
\begin{equation}
\label{eqn:LYscore}
U(\bbeta)
:=
\left[\sum_{i=1}^n 
\Delta_i  [z_i-\tilde{z}_n(t_i)]\right]
-
\left[\sum_{i=1}^n  \left( \int_0^{t_i}  [z_i-\tilde{z}_n(u)]^{\otimes 2}  du
\right)\right] 
\bbeta = 0,
\end{equation}
where for a given column vector $a$, we denote the matrix 
$a^{\otimes 2} = a \, a'$, and  
\begin{equation}
\tilde{z}_n(u):= \dfrac
{\sum_{i=1}^n z_i 1(t_i \geq u)}
{\sum_{i=1}^n 1(t_i \geq u)}
\end{equation}
is the vector function which averages
of all the $z$'s corresponding to time values greater or equal to $u$. \\

In order to put the estimating function $U(\bbeta)$ of Equation (\ref{eqn:LYscore}) 
into a Bayesian context, let us now 
denote the statistics
\begin{eqnarray*}
	V_1 &:=& \frac{1}{n} \sum_{i=1}^n 
	\Delta_i  [z_i-\tilde{z}_n(t_i)], 
	\\
	V_2 &:=& \frac{1}{n} 
	\sum_{i=1}^n  \left( \int_0^{t_i}  [z_i-\tilde{z}_n(u)]^{\otimes 2}  du
	\right),
	\\
	V_3 &:=& \frac{1}{n}
	\sum_{i=1}^n   [z_i-\tilde{z}_n(t_i)]^{\otimes 2}  
	,
\end{eqnarray*}
where $V_1$ is a row vector, while $V_2$ and $V_3$ are symmetric matrices. Notice that
with that notation LY estimating equation is $U(\bbeta)=V_1-V_2 \, \bbeta=0$. 
Consider further the function
\begin{equation}
g(\bbeta) = 
\exp \left\{
-\frac{1}{2}
\left[
(\bbeta - V_2^{-1}\, V_1)'
\left(\dfrac{V_2^{-1}\, V_3 \, V_2^{-1}}{n}\right)^{-1}    
(\bbeta - V_2^{-1}\, V_1)
\right]
\right\}
\label{eqn:mmimick}.
\end{equation}
Taking logarithms and derivatives with respect to $\bbeta$ we see that 
\begin{eqnarray*}
	\log g(\bbeta)\!\!\!
	& = & \!\!
	-\frac{1}{2}
	\left[
	\bbeta' \, \left(\dfrac{V_2^{-1}\, V_3 \, V_2^{-1}}{n}\right)^{-1}  \bbeta 
	- 2 \, \bbeta' \, 
	\left(\dfrac{V_2^{-1}\, V_3 \, V_2^{-1}}{n}\right)^{-1}
	\,
	(V_2^{-1}\, V_1) 
	%(V_2^{-1}\, V_3 \, V_2^{-1})^{-1} \,
	% \bbeta  
	\right.
	\\
	&& \left. \quad - (V_2^{-1}\, V_1)'
	\left(\dfrac{V_2^{-1}\, V_3 \, V_2^{-1}}{n}\right)^{-1}    
	(V_2^{-1}\, V_1)
	\right],
	\\
	\frac{\partial}
	{\partial \bbeta} \log g(\bbeta)\!\!\!
	&=&\!\!
	- \left(\dfrac{V_2^{-1}\, V_3 \, V_2^{-1}}{n}\right)^{-1}  \bbeta
	+
	\left(\dfrac{V_2^{-1}\, V_3 \, V_2^{-1}}{n}\right)^{-1}
	(V_2^{-1}\, V_1)  
	.
\end{eqnarray*}
From the display above, we observe that $U(\bbeta)=0$ whenever $(\partial/\partial \bbeta) \log g(\bbeta) = 0$. 
As a consequence, it is remarkable that Lin and Ying's point estimator 
of $\bbeta$ corresponds to the mean or the mode of a multivariate normal
distribution with mean vector $\mathbf{m}=(V_2^{-1}\, V_1)$ and covariance matrix 
is $\mathbf{D}=n^{-1}\,(V_2^{-1}\, V_3 \, V_2^{-1})$. This entails that in a Bayesian context 
we could consider LY estimators as belonging to a posterior normal distribution for
the Euclidean parameter with a flat (improper) prior.

In this manuscript, in contrast, we have opted for an informative conjugate multivariate normal prior 
truncated to the positive orthant given by Equation (\ref{eqn:prioribeta}).
From there, we propose the pseudo-marginal 
posterior distribution for  $[\mathbf{B}|\mathcal{X},\mathbf{L}_0]$ as
\begin{multline}
f _{\mathbf{B}|\mathcal{X},\mathbf{L}_0}
\big{(}\bbeta\big{)}
\propto 
\exp\left\{
-\frac{1}{2}
\left[
(\bbeta-\mathbf{m})'
\mathbf{D}^{-1}
(\bbeta-\mathbf{m})'
+
(\bbeta-\mmu_\beta)'
\mathbf{C}_\beta^{-1}
(\bbeta-\mmu_\beta)'
\right]
\right\}\\ 
\prod_{j=1}^k I(\beta_j \geq 0)
,
\label{eqn:postdensbeta}
\end{multline}
which from standard calculations entail
\begin{equation}
[\mathbf{B}|\mathcal{X},\mathbf{L}_0]
\sim 
N_k 
\left[
(\mathbf{D}^{-1} + \mathbf{C}_\beta^{-1})^{-1} (\mathbf{D}^{-1} \mathbf{m}+ \mathbf{C}_\beta^{-1} \mmu_\beta)
;
(\mathbf{D}^{-1} + \mathbf{C}_\beta^{-1})^{-1} 
\right] \Big{|}_{\mathbb{R}^{k+}}.
\label{eqn:postbeta}
\end{equation}
It is worthy to remark, at this point, why we call the distribution in the Equation above a ``\textit{pseudo}'' posterior, instead of a (plain) posterior distribution. That is because the Bayesian calculations 
that led to the density $f _{\mathbf{B}|\mathcal{X},\mathbf{L}_0}$ in Equation (\ref{eqn:postdensbeta})
started from $g(\bbeta)$, which is not the conditional distribution of the sample given the parameters,
but a convenient transformation inspired by an estimating equation. That is why we call our method 
``\textit{hybrid}''. Within a Bayesian framework, our method shares the same spirit of Cox's \textit{partial} likelihood, and Lin and Ying's estimating equation in an alternative frequentist approach to inference. All those methods rely on estimating equations that depart from the traditional likelihood function.
Furthermore, it turns out convenient that with our formulation
$[\mathbf{B}|\mathcal{X},\mathbf{L}_0]=
[\mathbf{B}|\mathcal{X}]$.
I.e., this marginal posterior distribution is 
independent of the baseline hazard function. This enables estimation of $\bbeta$ 
to be disentangled from that of $\lambda_0(\cdot)$, which in many contexts is treated as a
\textit{nuisance} parameter. Secondly, we will see that our formulation 
enables point estimation of $\bbeta$ in closed-form, a feature which is shared by LY development, 
while not by Cox's. 

As for the choice of the Bayesian point estimator of $\bbeta$,  we opt in this 
article for
the mode of the posterior distribution. This is,
\begin{equation}
\hat{\bbeta}^{\text{B}}
=
(\mathbf{D}^{-1} + \mathbf{C}_\beta^{-1})^{-1} (\mathbf{D}^{-1} \mathbf{m}+ \mathbf{C}_\beta^{-1} \mmu_\beta)
%\nonumber
%% \\
%% &=&
%% (V_2\, V_3^{-1} \, V_2 + \mathbf{C}_\beta^{-1})^{-1}
%% \,
%% (V_2\, V_3^{-1} \, V_1 
%% + \mathbf{C}_\beta^{-1} \mmu_\beta)
\label{eqn:postmodebeta}
\end{equation}
whenever positive, or zero otherwise. 

It is notheworthy that that under a noninformative prior,
i.e., when  \break
$\|\mathbf{C}_\beta\|\to \infty$, then the point estimator
$\hat{\bbeta}^{\text{B}}$ becomes asymptotically equivalent to $\mathbf{m} = V_2^{-1} \, V_1 
$,
which coincides in the positive case with LY classical estimator. In contrast with LY estimator,
however, ours has the advantage that it could never turn out negative for any sample. Notice also that 
an alternative Bayesian estimator,
based on the expectation of the pseudo-posterior distribution, would have a serious 
problem: that it would diverge to infinity as $\|\mathbf{C}_\beta\|\to \infty$, due to the effect of 
truncation of the prior multivariate normal distribution to the positive orthant. For the same reason,
we propose not to estimate the variance of $\hat{\bbeta}^{\text{B}}$ by the variance of the  
(truncated) pseudo posterior distribution, but instead with a different method. We explain that as 
follows. 

\paragraph{Estimation of the Variance.}
We proceed in the following steps
\begin{enumerate}
	\item Estimate $\hat{\bbeta}^\text{B}$ according to Equation (\ref{eqn:postmodebeta}) whenever
	positive, or zero otherwise.
	
	\item To estimate the precision of each component $\hat{\bbeta}_k^\text{B}$ we will construct 
	the highest posterior density (HPD) credible interval
	of $(1-\alpha)$ coverage.  The interval is of 
	the form $[b_l,b_u]$, where $0 \leq b_l < b_u < \infty$. It is noteworthy that in Lin and Ying's formulation, since the 
	confidence intervals are based on the approximate normal distribution, they take the 
	form
	$
	\hat{\beta}^\text{LY} \pm z_{1-\alpha/2} 
	\, \hat{\sigma}_{\left(\hat{\beta}^\text{LY}\right)}
	$,
	which allows for a possibly negative lower endpoint. 

	Here for the sake of comparability of the standard deviation estimators, we propose
	\begin{equation}
	\hat{\sigma}_{\left(\tilde{\beta}^\text{B}\right)}
	:= 
	\dfrac
	{b_u - b_l
	}
	{2 \, z_{1-\alpha/2}
	}
	.
	\end{equation}
	Notice also that because we work with truncated normal distributions, the point
	estimates need not be at the center of the intervals, which will happen only when $b_l>0$. 
\end{enumerate}

This method for estimating $\tilde{\sigma}_{\left(\tilde{\beta}^\text{B}\right)}$ also has
an important consequence for testing the significance of 
covariates. Based on the duality between confidence intervals and testing, we 
will consider that when zero belongs to an interval, it is an indication of 
nonsignificance of a particular covariate. 
In this regard, constructing credible intervals of highest pseudo-posterior
density is essential  for testing purposes as well as model selection.

\subsection{Estimation of the Baseline Hazard}\label{SS:EBH}

We recall first the so-called 
Gamma Process. 
Let $\mathcal{G}(a,b)$ denote the gamma distribution with shape parameter $a>0$ and scale parameter $b>0$. Let $\alpha(t)$ be an increasing left continuous function on $t \geq 0$ such that
$\alpha(0)=0$. Further, let $Z(t)$ be a stochastic process on $t\geq 0$, with the properties: 
(\textit{i}) $Z(0)=0$, (\textit{ii}) $Z(t)$ has independent increments in disjoint intervals, and
(\textit{iii})	For  any $t>s$, $Z(t) - Z(s) \thicksim \mathcal{G}(c(\alpha(t)-\alpha(s)),c)$.		
Then the process $\{Z(t):t>0\}$ is called a \textit{Gamma Process} and is denoted by $Z(t)\thicksim \mathcal{GP}(c\alpha(t),c)$.

We opt in this manuscript to incorporate a Gamma Process prior on the
cumulative baseline hazard function. With that goal, we first need to find an 
alternative expression for the likelihood of the AHM, expressed in terms of 
the cumulative hazard increments 
$\Lambda_0^+ = (\Lambda_{0_1}^+,\Lambda_{0_2}^+,\cdots,\Lambda_{0_m}^+:=\infty)$ 
associated to each interval in the grid. Since there is no sense in trying to estimate the cumulative increment in the last interval of the grid, we truncate it at a fixed time $t_F$ (i.e. $s_m=t_F$); this guarantees that $\Lambda_{0_m}^+<\infty$. This does not represent a limitation in practice, because every empirical study has necessarily a finite follow up time.\\ 
Considering the expression (\ref{eqn:steplambda}), which introduced the $A_j$'s, we deduce 
$\Lambda_{0_j}^+ = A_j  (s_j - s_{j-1})$. Then, replacing for each $j$ we obtain:
\begin{equation}
f_{\Lambda_{0_j}^+\vert \mathcal{X},\bbeta}(a_{0_j}^+) \propto  \sum_{k=0}^{N_{j}}d_{k}\left(\frac{a_{0_j}^+}{s_j\! -\! s_{j-1}}\right)^{k}e^{-\left(\frac{a_{0_j}^+}{s_j - s_{j-1}}\right)\left(\left(\sum_{k_j=1}^{n_j}(t_{k_j}- s_{j-1})\right)+ m_j(s_j - s_{j-1})\right)}.
\end{equation}
For a fixed and increasing left continuous function $\alpha$ and a scale parameter $c$, the prior for each $\Lambda_{0_j}^+$ is $\mathcal{G}(c(\alpha(s_j)-\alpha(s_{j-1})),c)$. By construction, different grid intervals are disjoint, and following the gamma process structure, they are independent. 
According to that, the prior for $\Lambda_{0}^+$ is:
\[\pi(\Lambda_{0}^+) = \prod_{j=1}^{m} f_{\mathcal{G}(c(\alpha(s_j)-\alpha(s_{j-1})),c)}(a_{0}^+) = \prod_{j=1}^{m} f_{\mathcal{G}(c(\alpha_j),c)}(a_{0}^+), \]
where $\alpha_j = \alpha(s_j)- \alpha(s_{j-1})$, and $f_{\mathcal{G}(a,b)}(t)$ denotes the density function of a Gamma distribution with parameters $a$ and $b$ evaluated at $t$. Using that factor, we obtain the posterior
\begin{multline*}
f_{\Lambda_{0}^+\vert \mathcal{X},\bbeta}(a_{0}^+) \!\propto\! \prod_{j=1}^{m}  \sum_{k=0}^{N_{j}}d_{k}\!\left(\frac{a_{0_j}^+}{s_j\! -\! s_{j-1}}\right)^{k}e^{-\left(\frac{a_{0_j}^+}{s_j \!-\! s_{j-1}}\right)\left(\left(\sum_{k_j=1}^{n_j}(t_{k_j}\!-\! s_{j-1})\right)+ m_j(s_j - s_{j-1})\right)}
\\
\prod_{j=1}^{m} \frac{(a_{0_j}^+)^{c\alpha_j -1}\,e^{-a_{0_j}^+\,c}\,c^{c\, \alpha_j}}{\Gamma(c \, \alpha_j)}.
\end{multline*}
In order now to simplify notation, let us call
\begin{align*}
d_k^{(j)} &:= \frac{d_k}{(s_j - s_{j-1})^k \times \Gamma(c \, \alpha_j)}, \\
c_j &:=\frac{\left(\left(\sum_{k_j=1}^{n_j}(t_{k_j}- s_{j-1})\right)+ m_j(s_j - s_{j-1})\right)}{(s_j - s_{j-1})} + c .
\end{align*}
With that notation we express the posterior
\begin{eqnarray*}
	f_{\Lambda_{0_j}^+\vert \mathcal{X},\bbeta}(a_{0}^+) &=&
	\frac{\sum_{k=0}^{N_j}d_k^{(j)} \, (a_{0_j}^+)^{\left(k+ c \, \alpha_j - 1 \right)} e^{-a_{0_j}^+ \, c_j}}
	{\int_{-\infty}^{\infty}\sum_{k=0}^{N_j}d_k^{(j)} \, (a_{0_j}^+)^{\left(k+ c.\alpha_j - 1 \right)} e^{-a_{0_j}^+ . c_j}da_{0}^+}
	\\
	%% &=\frac{\sum_{k=0}^{N_j}d_k^{(j)}.(a_{0_j}^+)^{\left(k+ c.\alpha_j - 1 \right)} e^{-a_{0_j}^+ . c_j}}{\sum_{k=0}^{N_j}d_k^{(j)}\int_{-\infty}^{\infty}(a_{0_j}^+)^{\left(k+ c.\alpha_j - 1 \right)} e^{-a_{0_j}^+ . c_j}da_{0}^+}\\
	%% &=\frac{\sum_{k=0}^{N_j}d_k^{(j)}.(a_{0_j}^+)^{\left(k+ c.\alpha_j - 1 \right)} e^{-a_{0_j}^+ . c_j}}{\sum_{k=0}^{N_j}d_k^{(j)}\int_{-\infty}^{\infty}(a_{0_j}^+)^{\left(k+ c.\alpha_j - 1 \right)} e^{-a_{0_j}^+ . c_j}da_{0}^+}\\
	%
	&=&
	\frac{\sum_{k=0}^{N_j}\frac{d_k^{(j)}}{c_j^k} \, \Gamma(k+c \, \alpha_j) \, f_{\mathcal{G}(k+c \, \alpha_j,c_j)}(a_{0}^+)}
	{\sum_{k=0}^{N_j}\frac{d_k^{(j)}}{c_j^k}\, \Gamma(k+c\,\alpha_j)}.
\end{eqnarray*}
Therefore, the resulting posterior distribution is proportional to a
mixture of Gamma distributions, which enables point estimation in closed-form under square error loss. This is one of the most important advantages of our choice of prior specification in this article.
Naturally, other 
choices of prior distributions for the baseline hazards are possible, for example by placing a prior 
on the plain hazard function itself, or on the baseline survival function. Those options will be presented in a separated article, a first version of which could be obtained from the authors upon request.

In order now to provide formulae for the point estimators and their variances, let us call
$e_k^{(j)} :=({d_k^{(j)}}/{c_j^k}) \, \Gamma(k+c \, \alpha_j)$, so that
\begin{eqnarray*}
	\widehat{a_{0_j}^+}^B
	=
	\text{E}[\Lambda_{0_j}^+\vert \mathcal{X},\bbeta]
	&=&
	\int_{-\infty}^{\infty}a_{0}^+ \, \left[\dfrac{\sum_{k=0}^{N_j}e_k^{(j)}.f_{\mathcal{G}(k+c.\alpha_j,c_j)}(a_{0}^+)}{\sum_{k=0}^{N_j}e_k^{(j)}}\right] da_{0}^+
	\\
	%% &=\frac{\sum_{k=0}^{N_j}e_k^{(j)}.\int_{-\infty}^{\infty}a_{0}^+.[f_{\mathcal{G}(k+c.\alpha_j,c_j)}(a_{0}^+)]da_{0}^+}{\sum_{k=0}^{N_j}e_k^{(j)}} \\
	%
	&=&
	\dfrac{\sum_{k=0}^{N_j}e_k^{(j)} \, (k+c \, \alpha_j)}{\sum_{k=0}^{N_j}e_k^{(j)}\, c_j}.
\end{eqnarray*}
To estimate their variances, we take
\begin{eqnarray*}
	V[\Lambda_{0_j}^+\vert \mathcal{X},\bbeta]
	&=&
	\left(\int_{-\infty}^{\infty}(a_{0}^+)^2\, \left[\frac{\sum_{k=0}^{N_j}e_k^{(j)} \, 
		f_{\mathcal{G}(k+c.\alpha_j,c_j)}(a_{0}^+)}
	{\sum_{k=0}^{N_j}e_k^{(j)}}\right] da_{0}^+\right)
	-E[\Lambda_{0_j}^+\vert \mathcal{X},\bbeta]^2 
	\\
	%% &=\left(\frac{\sum_{k=0}^{N_j}e_k^{(j)}.\int_{-\infty}^{\infty}(a_{0}^+)^2.[f_{\mathcal{G}(k+c.\alpha_j,c_j)}(a_{0}^+)]da_{0}^+}{\sum_{k=0}^{N_j}e_k^{(j)}}\right) - (\widehat{a_{0_j}^+}^B)^2 \\
	&=&
	\dfrac{\sum_{k=0}^{N_j}e_k^{(j)} \, [(k+c \, \alpha_j) \,  c_j^2+((k+c \, \alpha_j) \, c_j)^2]}
	{\sum_{k=0}^{N_j}e_k^{(j)}} - (\widehat{a_{0_j}^+}^B)^2.
\end{eqnarray*}

As with our proposed the estimator for $\bbeta$, it is worth to note the properties of 
$\widehat{a_{0_j}^+}^\text{B}$ as we relax or tighten the prior.

\textbf{Remark 1:} When $c$ diverges to $\infty$, each $\widehat{a_{0_j}^+}^B$ 
converges almost surely to $\alpha_j$ (i.e., it's prior mean). 
To see that, we express 
\begin{eqnarray*}
	\widehat{a_{0_j}^+}^B &=&
	\frac{\sum_{k=0}^{N_j}e_k^{(j)} \, (k+c \, \alpha_j)}
	{\sum_{k=0}^{N_j}e_k^{(j)} \,c_j} 
	=
	\frac{\sum_{k=0}^{N_j}e_k^{(j)}\,k}
	{\sum_{k=0}^{N_j}e_k^{(j)} \,c_j} 
	+
	\frac{\sum_{k=0}^{N_j}e_k^{(j)} \, c \, \alpha_j}
	{\sum_{k=0}^{N_j}e_k^{(j)} \,c_j} 
	\\
	&=&
	\frac{\sum_{k=1}^{N_j}e_k^{(j)} \, k}{e_0^{(j)} \, c_j+\sum_{k=1}^{N_j}e_k^{(j)} \, c_j}
	+ \frac{c \,\alpha_j}{c_j}. 
\end{eqnarray*}
Since $e_k^{(j)}>0$ for all $k \in \{0,1,\cdots,N_j\}$, 
\[
0 \leq \frac{\sum_{k=1}^{N_j}e_k^{(j)} \,k}
{e_0^{(j)} \, c_j + 
	\sum_{k=1}^{N_j}e_k^{(j)} \, c_j}  \leq 
\frac{\sum_{k=1}^{N_j}e_k^{(j)} \, N_j}
{\sum_{k=1}^{N_j} e_k^{(j)} \, c_j} 
=\frac{N_j}{c_j}.
\]
Since $N_j$ is bounded in probability, $\widehat{a_{0_j}^+}^\text{B}$ converges in probability to 
$\alpha_j$ as $c \rightarrow \infty$.

\textbf{Remark 2:} In the opposite case than the first Remark, i.e., when
$c$ converges to zero, the Bayesian estimator, $\hat{a}_{0_j}^{+\text{B}}$ does not depend 
on the prior function $\alpha_j$. The proof of that is clear since every time that $\alpha_j$
is in the expression, it is appears multiplied by $c$.

It is worth to mention again that our choices of the  loss functions is different for the 
Euclidean coefficients than for the baseline hazard function.  While we used the 
mode of the posterior distribution for estimation of the Euclidean parameter $\bbeta$, we have 
opted to use the mean of the posterior distribution for estimation of the baseline hazard. It is also for this reason that we call at our method is \textit{hybrid}.
Again, we notice that one of the main advantages of this approach is that it enables estimators in closed form. 

\section{Simulation}
\label{sec:simul}

In this Section, we present a small simulation experiment according to the following choices:
(\textit{i}) hazard function $\lambda(t,\beta) = 1 + 0.5 z$, i.e. $\lambda_0(t)=1$ in $\mathbb{R}^+$, and a single covariate with regression parameter $\beta=0.5$, (\textit{ii}) Censoring variable 
$C$ with an Exponential distribution with mean 2, 
(\textit{iii}) Covariate $Z \sim \chi^2_1$, (\textit{iv}) sample sizes $n=100$ or $500$,
and (\textit{v}) $R=1000$ replicates.

\paragraph{Simulations for the regression parameter $\beta$.} 
The results are presented in Tables (\ref{tabla1}) and (\ref{tabla2}). We observe that in all cases
$\hat{\beta}^\text{B} \geq 0$, and further, if either $\|\mathbf{C}_{\mathbf{\beta}}\| \to \infty$ 
(i.e., the prior becomes less informative)  
or $n \to \infty$ (the sample size increases), the point estimator of $\beta$ and its standard deviation converge to those of Lin 
and Ying, i.e., $\hat{\beta}^\text{B}  \approx \hat{\beta}^\text{LY}$ and 
$\hat{\sigma}^\text{B}(\hat{\beta}^\text{B}) \approx \hat{\sigma}^\text{LY}(\hat{\beta}^\text{LY})$.
Those simulation tables also show the effects of varying the position parameter $\mu_{\beta}$ 
and the scale parameter $w$ of the $\beta$-prior. Naturally, the estimates of $\beta$ get closer to
its true value ($0.5$) as the prior position parameter $\mu_{\beta}$ gets near to the truth.

{\small{
	\begin{table}[!h]
		\centering {
			%\small
			\begin{tabular}{c|ccccccc}\hline
				& & & &  $w$ & & &  \\ 
				$\mu_\beta$   &	  0.1   &     0.5   &     1     &     2     &     5     &     10    &    1000  \\ \hline
				0    & 0.3241 & 0.4813 & 0.5155 & 0.5352 & 0.5479 & 0.5524 & 0.5569 \\
				& (\textit{0.1625}) & (\textit{0.2122}) & (\textit{0.2211}) & (\textit{0.2259}) & (\textit{0.2291}) & (\textit{0.2301}) & (\textit{0.2312}) \\ \hline
				0.25 & 0.4157 & 0.5093 & 0.5306 & 0.5430 & 0.5512 & 0.5540 & 0.5569 \\
				& (\textit{0.1784}) & (\textit{0.2154}) & (\textit{0.2227}) & (\textit{0.2268}) & (\textit{0.2294}) & (\textit{0.2303}) & (\textit{0.2312}) \\ \hline
				0.5  & 0.5073 & 0.5372 & 0.5457 & 0.5509 & 0.5544 & 0.5556 & 0.5569 \\
				& (\textit{0.1851}) & (\textit{0.2182}) & (\textit{0.2242}) & (\textit{0.2275}) & (\textit{0.2297}) & (\textit{0.2304}) & (\textit{0.2312}) \\ \hline
				0.75 & 0.5989 & 0.5652 & 0.5608 & 0.5587 & 0.5576 & 0.5573 & 0.5569 \\
				& (\textit{0.1872}) & (\textit{0.2203}) & (\textit{0.2256}) & (\textit{0.2283}) & (\textit{0.2300}) & (\textit{0.2306}) & (\textit{0.2312}) \\ \hline
				1    & 0.6905 & 0.5932 & 0.5759 & 0.5666 & 0.5608 & 0.5589 & 0.5569 \\
				& (\textit{0.1879}) & (\textit{0.2221}) & (\textit{0.2268}) & (\textit{0.2290}) & (\textit{0.2303}) & (\textit{0.2307}) & (\textit{0.2312}) \\ \hline
				2    & 1.0570 & 0.7051 & 0.6362 & 0.5981 & 0.5737 & 0.5654 & 0.5570 \\
				& (\textit{0.1884}) & (\textit{0.2262}) & (\textit{0.2304}) & (\textit{0.2314}) & (\textit{0.2315}) & (\textit{0.2313}) & (\textit{0.2312}) \\ \hline
				10   & 3.9886 & 1.6002 & 1.1189 & 0.8496 & 0.6770 & 0.6175 & 0.5575 \\
				& (\textit{0.1884}) & (\textit{0.2290}) & (\textit{0.2365}) & (\textit{0.2394}) & (\textit{0.2378}) & (\textit{0.2354}) & (\textit{0.2312}) \\ \hline
		\end{tabular}}
		\caption{
		%\small{}
		Simulations for the regression parameter $\beta$:	$n=100$, LY = 0.5569 (\textit{0.2457}). }
	%}
	\label{tabla1}
\end{table}}}

{\small{
	\begin{table}[!h]
			\centering {	
			\begin{tabular}{c|ccccccc}\hline
				%			$\sigma_\beta$   &
				%			$\mu_\beta=0$ &
				%			$\mu_\beta=0.25$   &
				%			$\mu_\beta=0.5$ &
				%			$\mu_\beta=0.75$ &
				%			$\mu_\beta=1$ &
				%			$\mu_\beta=2$ &
				%			$\mu_\beta=10$ \\ 
				& & & &  $w$ & & &  \\ 
				$\mu_\beta$   &	  0.1   &     0.5   &     1     &     2     &     5     &     10    &    1000\\ \hline
				0    &   0.4585 & 0.4992   & 0.5048   & 0.5077   & 0.5094   & 0.5100   & 0.5106   \\
				& (\textit{0.0982}) & (\textit{0.1024}) & (\textit{0.1030}) & (\textit{0.1033}) & (\textit{0.1035}) & (\textit{0.1035}) & (\textit{0.1036}) \\ \hline
				0.25 & 0.4829 & 0.5045 & 0.5075 & 0.5090 & 0.5100 & 0.5103 & 0.5106 \\
				& (\textit{0.0982}) & (\textit{0.1024}) & (\textit{0.1030}) & (\textit{0.1033}) & (\textit{0.1035}) & (\textit{0.1035}) & (\textit{0.1036}) \\ \hline
				0.5  & 0.5074 & 0.5099 & 0.5102 & 0.5104 & 0.5105 & 0.5106 & 0.5106 \\
				& (\textit{0.0982}) & (\textit{0.1024}) & (\textit{0.1030}) & (\textit{0.1033}) & (\textit{0.1035}) & (\textit{0.1035}) & (\textit{0.1036}) \\ \hline
				0.75 &   0.5319 & 0.5152 & 0.5129 & 0.5118 & 0.5111 & 0.5108 & 0.5106 \\
				& (\textit{0.0982}) & (\textit{0.1024}) & (\textit{0.1030}) & (\textit{0.1033}) & (\textit{0.1035}) & (\textit{0.1035}) & (\textit{0.1036}) \\ \hline
				1    &   0.5564 & 0.5205 & 0.5156 & 0.5131 & 0.5116 & 0.5111 & 0.5106 \\
				& (\textit{0.0982}) & (\textit{0.1024}) & (\textit{0.1030}) & (\textit{0.1033}) & (\textit{0.1035}) & (\textit{0.1035}) & (\textit{0.1036}) \\ \hline
				2    &   0.6543 & 0.5419 & 0.5264 & 0.5186 & 0.5138 & 0.5122 & 0.5106  \\
				& (\textit{0.0982}) & (\textit{0.1024}) & (\textit{0.1030}) & (\textit{0.1033}) & (\textit{0.1035}) & (\textit{0.1035}) & (\textit{0.1036}) \\ \hline
				10   &   1.4375 & 0.7128 & 0.6128 & 0.5620 & 0.5312 & 0.5209 & 0.5107 \\
				& (\textit{0.0982}) & (\textit{0.1024}) & (\textit{0.1030}) & (\textit{0.1033}) & (\textit{0.1035}) & (\textit{0.1035}) & (\textit{0.1036}) \\ \hline
		\end{tabular}}
		\caption{{Simulations for the regression parameter $\beta$: $n=500$, LY = 0.5106 (\textit{0.1036}). }}\label{tabla2}
\end{table}}} 

\paragraph{Simulation for the baseline hazard parameters.}

In this second part of the simulation experiment, we choose a grid 
according the approximate quantiles $0.2, 0.4, 0.6$ and $0.8$ of the observed event times.
A decreasing number of the ``at risk'' times along the grid makes more accurate those 
estimates in earlier intervals in the time axis, as is seen in estimates of 
the standard deviations. According to the chosen grid, the true values are 
$\Lambda_0^+ = (0.125, 0.3, 0.6, 1.15)$. 
Also, for this part of the simulation, the priori selected is a left continuous 
function $\alpha(t)$, which takes values:
$\alpha(0)=0$, $\alpha(0.125)=5$, $\alpha(0.3)=6$, $\alpha(0.6)=6.3$, and $\alpha(1.15)=6.31$.
Thus, the corresponding true values of the vector 
$(\alpha_1,\alpha_2,\alpha_3,\alpha_4)=(5,1,0.3,0.01)$. 
The prior weight is regulated by the hyperparameter $c$. We present the simulation with $c=10$, $c=1$ and $c=0.1$.
Also, for this simulation, we have previously estimated $\widehat{\bbeta}$ with the hybrid Bayesian estimator, with prior parameters $\mu_{\bbeta}=0.5$ and $\sigma_{\bbeta}=10000$.

The results are presented in Tables (\ref{tabla3}) and (\ref{tabla4}).
As expected, we see that (\textit{i}) the value of $c$ has an impact in the estimator,
giving more or less weight to the prior,
(\textit{ii}) the standard deviation increases as the intervals shift to the right, 
and (\textit{iii}) as the sample size increases, the prior losses impact on the estimates.

\begin{table}
\begin{center}
	\begin{tabular}{lllll}
		\hline\noalign{\smallskip}
		c & $\Lambda_{0_1}^+$=0.125 & $\Lambda^+_{0_2}$=0.175 & $\Lambda_{0_3}^+$=0.3 & $\Lambda_{0_4}^+$=0.55  \\
		\noalign{\smallskip}\hline\noalign{\smallskip}
		$10$   & 0.6527 & 0.2967 & 0.2991 & 0.3354\\     
		& (\textit{0.0824}) & (\textit{0.0657}) & (\textit{0.0825}) & (\textit{0.1127})  \\
		$1$  & 0.1880 & 0.1845 & 0.2967 & 0.5167 \\
		& (\textit{0.0495}) & (\textit{0.0587}) &  (\textit{0.0928}) &  (\textit{0.1762})  \\
		$0.1$   & 0.1265 & 0.1688 & 0.2951 & 0.5384\\
		& (\textit{0.0431}) & (\textit{0.0577}) & (\textit{0.0943}) & (\textit{0.1850})    \\
		\noalign{\smallskip}\hline
	\end{tabular}
\end{center}	
\caption{Simulation for the baseline hazard parameters: $n=100$, $\widehat{\bbeta} = 0.5569$.}
\label{tabla3}       % Give a unique label
\end{table}

	\begin{table}
\begin{center}
	\begin{tabular}{lllll}
		\hline\noalign{\smallskip}
		c & $\Lambda_{0_1}^+$=0.125 & $\Lambda^+_{0_2}$=0.175 & $\Lambda_{0_3}^+$=0.3 & $\Lambda_{0_4}^+$=0.55  \\
		\noalign{\smallskip}\hline\noalign{\smallskip}
		$10$   & 0.2510 & 0.2041 & 0.2991 & 0.4919 \\
		& (\textit{0.0247}) & (\textit{0.0269}) & (\textit{0.0406}) &  (\textit{0.0732})\\
		$1$   & 0.1384 & 0.1771 & 0.2990 & 0.5437 \\
		& (\textit{0.0198}) & (\textit{0.0260}) & (\textit{0.0416}) & (\textit{0.0799}) \\
		$0.1$   & 0.1257 & 0.1742 & 0.2990 & 0.5494\\
		& (\textit{0.0192}) & (\textit{0.0259}) &  (\textit{0.0417}) &  (\textit{0.0807})  \\
		\noalign{\smallskip}\hline
	\end{tabular}\end{center}
\caption{Simulation for the baseline hazard parameters: $n=500$, $\widehat{\bbeta} = 0.5106$.}
\label{tabla4}       % Give a unique label

\end{table}

\section{Analysis of a real dataset}
\label{sec:data}
In this last Section we compute and compare the hybrid Bayesian estimates of $\bbeta$ 
from the Welsh Nickels
miners dataset, with those of Lin and Ying (1994). \\
The ``Welsh Nickels refinery'' dataset was originally introduced by Doll \textit{et al.} (1970), and subsequently analyzed by Breslow \& Day (1987), Lin \& Yin (1994) and \'{A}lvarez \& Ferrario (2016), among others. It contains information about 679 workers from a nickel refinery at the south of Wales, who were presumably exposed to cancer triggering substances. Data were collected from payroll registers from 1934 to 1981. The covariates are AFE:= Age at first employment, YFE:= Year at first employment and EXP:= Exposure level (an index of the degree of contamination). We fit the same model as in Lin \& Ying (1994). We observed that the hybrid Bayesian estimate approaches the classical LY estimator as the prior gets flatter.
%
%\begin{table}[!h]
%	\caption{\small{}}\label{tabla5}
%	\centering {\small
%		\begin{tabular}{c|cc|cc|cc}\hline
%			&\;\;\;\;\;\;\;\;\;\;\;\;\;L&Y\;\;\;\;\;\;\;\;\;\;\;\;\; & $\mu_{\bbeta}=0$ & $\sigma_{\bbeta}=1$ &  $\mu_{\bbeta}=0$ & $%			&
%			$\hat{\beta}_{LY}$   &
%			$\hat{\sigma}_{\hat{\beta}_{LY}}$ &
%			$\hat{\beta}$   &
%			$\hat{\sigma}_{\hat{\beta}}$ &
%			$\hat{\beta}$   &
%			$\hat{\sigma}_{\hat{\beta}}$ \\		
%			\hline
%			log(AFE-10) & 0.0025126 &  \textit{0.0004684} &  0.00254579 &    \textit{0.00143274}   &     0.0025126   &    \textit{0.0014250}  \\
%			(YFE-1915)/10 & 0.0002179 &  \textit{0.0003756} &  0.00021872 &    \textit{0.00087757}   &     0.0002179   &    \textit{0.0008774}  \\
		%-(YFE-1915)$^2/100$ & 0.0026005 &  \textit{0.0008819} &  0.00260866 &    \textit{0.00196908}   &     0.0026005   &    \textit{0.0019673}  \\
%			log(EXP+1) & 0.0017172 &  \textit{0.0004348} &  0.00171391 &    \textit{0.00097031}   &     0.0017172   &    \textit{0.0009711}  \\
			
%			\hline
%	\end{tabular}}
%\end{table}

\begin{table}[!h]
	%\caption{\small{}}
	\label{tabla5}
	\centering {\small
%\begin{table}
%\centering{\scalebox{0.87}{%
		\begin{tabular}{c|c|c|c|}%\hline
			\cline{2-4}& \multicolumn{1}{c|}{$\hat{\bbeta}_{LY}$} & $\mu_{\bbeta}=0$  $\sigma_{\bbeta}=0.01$ &  $\mu_{\bbeta}=0$  $\sigma_{\bbeta}=1000$ \\
			%\cline{2-7}&$\hat{\bbeta}_{LY}$  &$\hat{\sigma}_{\hat{\bbeta}_{LY}}$ &	$\hat{\bbeta}$   &	$\hat{\sigma}_{\hat{\bbeta}}$ &	$\hat{\bbeta}$   &
			%$\hat{\sigma}_{\hat{\bbeta}}$ \\		
			\hline
			\multicolumn{1}{|c|}{log(AFE-10)} & 2.5126 &  2.4183 &  2.5126   \\ %$\dfrac{\text{YFE-1915}}{10}$
			\multicolumn{1}{|c|}{(YFE-1915)/10} & 0.2179 &   0.1868 &   0.2179  \\ %$\dfrac{-(\text{YFE-1915})^2}{100}$ 
			\multicolumn{1}{|c|}{-(YFE-1915)$^2$/100}& 2.6005 &   2.3829 &  2.6005   \\ 
			\multicolumn{1}{|c|}{log(EXP+1)} & 1.7172 &   1.7081 &  1.7172   \\ 	\hline
\end{tabular}}
\caption{\small{(Results are multiplied by $10^3$)}}
\end{table}

The results are presented in Table (\ref{tabla5}), where we visualize the impact of the choice 
of the hyperparameters on the final estimates. Those results show the versatility of our proposed Hybrid Bayesian method, which can be calibrated to give the desired weight to the prior information. As discussed, this makes the
Hybrid Bayesian method a potential valuable tool for the Statistical practitioner in Data Analysis. 

\section{Conclusions}\label{sec:conc}
In this manuscript, we presented an alternative way to express the likelihood function of the AHM as a mixture of Gamma distributions. This enabled estimation methods in two steps: (i) estimation of the Euclidean coefficients in closed form, (ii) estimation of the baseline hazard function parameters. Our Bayesian proposal is based on constant regression coefficients, as did the work of LY. A flexible extension of those methods is to consider the general Aalen (1980) model, where $\beta=\beta(t)$, as in Silva \& Amaral-Turkman (2005), who additionally included frailty terms in their approach. \\

\vspace{0.4cm} \noindent{\bf   Acknowledgements}\\
	Maximiliano L. Riddick wishes to thank CONICET for his doctoral fellowship, and to Departamento de Matem\'{a}tica = Facultad de Ciencias Exactas - Universidad Nacional de La Plata.  Both authors thank Universidad Nacional de La Plata  (PPID UNLP I231).
%\end{acknowledgements}



%\begin{description}
\vspace{0.4cm} \noindent{\bf   References}\\
\begin{description}
	\item Aalen, O. (1980). A Model for Nonparametric Regression Analysis of Counting Processes. \textit{Mathematical statistics and probability theory}, 1-25.
	\item \'{A}lvarez, E. E. and Ferrario, J. (2016). Robust estimation in the additive hazards model. \textit{Communications in Statistics-Theory and Methods} \textbf{45 (4)}, 906-921.
	\item \'{A}lvarez, E. E. and Riddick, M. L. (2019). Review of Bayesian Analysis in Additive Hazards Model. \textit{Asian Journal of Probability and Statistics}, 1-14.
	\item \'{A}lvarez, E. E. and Ferrario, J. (2012). Revisi\'{o}n de la Estimaci\'{o}n Robusta en Modelos Semiparam\'{e}tricos de Supervivencia. \textit{IASI (Journal of the Interamerican Statistical Institute)} \textbf{64 (182-183)}, 85-106.
	\item Andersen, P. K., Borgan, O., Gill, R. D. and Keiding, N. (1993). \textit{Statistical Models Based on Counting Processes}. Springer. 
	\item Breslow, N. E. and Day, N. E. (1987). \textit{Statistical Methods in Cancer Research: The Design and Analysis of Cohort Studies}. International Agency for Research on Cancer.
	\item Cox, D. R. (1972). Regression Models and Life-Tables. \textit{Journal of the Royal Statistical Society. Series B (Methodological)} \textbf{34 (2)}, 187-220.
	\item Chernoukhov, A. (2013). Bayesian Spatial Additive Hazard Model. \textit{Electronic Theses and Dissertations. Windsor University.}
	\item  Chernoukhov, A., Hussein, A., Nkurunziza, S. and Bandyopadhyay, D. (2018). Bayesian inference in time-varying additive hazards models with applications to disease mapping. \textit{Environmetrics} \textbf{29 (5-6)}, e2478.
	\item Doll, R., Morgan, L. G. and Speizer, F. E. (1970). Cancers of the lung and nasal sinuses in nickel workers. \textit{British journal of cancer} \textbf{24 (4)}, 623-632.
    \item Dunson, D. B., \& Herring, A. H. (2005). Bayesian model selection and averaging in additive and proportional hazards models. Lifetime data analysis, 11, 213-232.
	\item Ibrahim, J. G., Chen, M. and Sinha, D. (2001). \textit{Bayesian Survival Analysis}. Springer Science \& Business Media.
	\item Kalbfleisch, J. D. and Prentice, R. L. (1990). \textit{The Statistical Analysis of Time Failure Data}. Hoboken: John Wiley \& Sons. 
    \item Klein, J. P. and Moeschberger, M. L. (2006). \textit{Survival analysis: techniques for censored and truncated data}. Springer Science \& Business Media.
	\item Lawless, J. F. (2003). Event history analysis and longitudinal surveys. \textit{Analysis of Survey data}, 221-243.
	\item Lin, D. Y. and Ying, Z. (1994). Semiparametric Analysis of the Additive Risk model. \textit{Biometrika} \textbf{81 (1)}, 61-71.
	\item Riddick, M. L. (2020). Estimaci\'on Bayesiana en el modelo de riesgos aditivos Doctoral dissertation, Universidad Nacional de La Plata, Argentina (https://doi.org/10.35537/10915/138368).
    \item Silva, G. L., \& Amaral-Turkman, M. A. (2005). Bayesian analysis of an additive survival model with frailty. Communications in Statistics-Theory and Methods, 33(10), 2517-2533.
    \item Turkman, M. A. A., Paulino, C. D., \& M\"uller, P. (2019). Computational bayesian statistics: an introduction (Vol. 11). Cambridge University Press.
	\item Zhang, J.  and  Lawson, A. B. (2011). Bayesian parametric accelerated failure time spatial model and its application to prostate cancer. \textit{Journal of Applied Statistics} \textbf{38 (3)}, 591-603.
\end{description}
\end{document}